\documentclass[11pt]{article}
\usepackage[utf8]{inputenc}
\usepackage{microtype}
\usepackage{lineno}  
\usepackage{xspace} 
\usepackage{caption} 

\usepackage{graphicx}  
\usepackage{color}
\usepackage{colortbl}
\graphicspath{{./figs/}} 
\usepackage{pdflscape}

\usepackage{siunitx} 
\usepackage{amsmath} 
\usepackage{amssymb}
\usepackage{todonotes}
\usepackage{amsfonts}
\usepackage{upgreek} 

\usepackage{jinstpub}
\usepackage{cleveref}

\usepackage{booktabs}
\usepackage{natbib}
\usepackage{graphicx}
\usepackage{siunitx}
\usepackage{hhline}
\usepackage{collcell}
\usepackage{tabularx}
\usepackage{comment}
\usepackage{makecell}
\usepackage{subcaption}
\usepackage{tikz}
\usepackage{tikz-3dplot}
\usetikzlibrary{arrows.meta}
\usepackage[nobiblatex]{xurl}

\usepackage{hyperref}
\hypersetup{%
  colorlinks=true,
  urlcolor=blue,
  linkcolor=blue,
  citecolor=blue
}

\usepackage{ifthen} 
\newboolean{pdflatex}
\setboolean{pdflatex}{true} 

\newboolean{articletitles}
\setboolean{articletitles}{true} 

\newboolean{uprightparticles}
\setboolean{uprightparticles}{false} 

\newboolean{inbibliography}
\setboolean{inbibliography}{false} 

\usepackage{xspace} 
\usepackage{upgreek}

\def\lhcb   {\mbox{LHCb}\xspace}

\def\cern {\mbox{CERN}\xspace}

\def\MagUp {\mbox{\em Mag\kern -0.05em Up}\xspace}

\ifthenelse{\boolean{uprightparticles}}%
{

 \def\PDelta      {\ensuremath{\Delta}\xspace}                 
 \def\PXi         {\ensuremath{\Xi}\xspace}                 
 \def\PLambda     {\ensuremath{\Lambda}\xspace}                 
 \def\PSigma      {\ensuremath{\Sigma}\xspace}                 
 \def\POmega      {\ensuremath{\Omega}\xspace}                 
 \def\PUpsilon    {\ensuremath{\Upsilon}\xspace}
 \let\oldPi\Pi
 \def\PPi         {\ensuremath{\oldPi}\xspace}

 \def\PB      {\ensuremath{\mathrm{B}}\xspace}                 
                  
 \def\PD      {\ensuremath{\mathrm{D}}\xspace}

 \def\PK      {\ensuremath{\mathrm{K}}\xspace}

 \def\Pi      {\ensuremath{\mathrm{i}}\xspace}

 \def\Ps      {\ensuremath{\mathrm{s}}\xspace}

 \def\thebaroffset{0.0em}
}
{

 \mathchardef\PDelta="7101
 \mathchardef\PXi="7104
 \mathchardef\PLambda="7103
 \mathchardef\PSigma="7106
 \mathchardef\POmega="710A
 \mathchardef\PUpsilon="7107
 \mathchardef\PPi="7105
                  
 \def\PB      {\ensuremath{B}\xspace}                 
                  
 \def\PD      {\ensuremath{D}\xspace}

 \def\PK      {\ensuremath{K}\xspace}

 \def\Pi      {\ensuremath{i}\xspace}

 \def\Ps      {\ensuremath{s}\xspace}

 \def\thebaroffset{0.18em}
}
\newcommand{\offsetoverline}[2][\thebaroffset]{\kern #1\overline{\kern -#1 #2}}%

\makeatletter
\ifcase \@ptsize \relax
  \newcommand{\miniscule}{\@setfontsize\miniscule{4}{5}}
\or
  \newcommand{\miniscule}{\@setfontsize\miniscule{5}{6}}
\or
  \newcommand{\miniscule}{\@setfontsize\miniscule{5}{6}}
\fi
\makeatother

\DeclareRobustCommand{\optbar}[1]{\shortstack{{\miniscule (\rule[.5ex]{1.25em}{.18mm})}
  \\ [-.7ex] $#1$}}

\def\squark    {{\ensuremath{\Ps}}\xspace}

\def\KorKbar {\kern \thebaroffset\optbar{\kern -\thebaroffset \PK}{}\xspace}

\def\D       {{\ensuremath{\PD}}\xspace}

\def\DorDbar {\kern \thebaroffset\optbar{\kern -\thebaroffset \PD}\xspace}

\def\Dp      {{\ensuremath{\D^+}}\xspace}
\def\Dm      {{\ensuremath{\D^-}}\xspace}

\def\DpDm    {\ensuremath{\Dp {\kern -0.16em \Dm}}\xspace}

\def\B       {{\ensuremath{\PB}}\xspace}

\def\BorBbar {\kern \thebaroffset\optbar{\kern -\thebaroffset \PB}\xspace}

\def\Bd      {{\ensuremath{\B^0}}\xspace}

\def\BdorBdbar {\kern \thebaroffset\optbar{\kern -\thebaroffset \Bd}\xspace}

\def\Bs      {{\ensuremath{\B^0_\squark}}\xspace}

\def\BsorBsbar {\kern \thebaroffset\optbar{\kern -\thebaroffset \Bs}\xspace}

\def\Y#1S{\ensuremath{\PUpsilon{(#1S)}}\xspace}

\def\LorLbar     {\kern \thebaroffset\optbar{\kern -\thebaroffset \PLambda}\xspace}

\def\AT#1     {\ensuremath{A_{\mathrm{T}}^{#1}}\xspace}           

\def\C#1      {\ensuremath{\mathcal{C}_{#1}}\xspace}                       
\def\Cp#1     {\ensuremath{\mathcal{C}_{#1}^{'}}\xspace}                    
\def\Ceff#1   {\ensuremath{\mathcal{C}_{#1}^{\mathrm{(eff)}}}\xspace}        
\def\Cpeff#1  {\ensuremath{\mathcal{C}_{#1}^{'\mathrm{(eff)}}}\xspace}       
\def\Ope#1    {\ensuremath{\mathcal{O}_{#1}}\xspace}                       
\def\Opep#1   {\ensuremath{\mathcal{O}_{#1}^{'}}\xspace}                    

\newcommand{\aunit}[1]{\ensuremath{\text{\,#1}}}       

\newcommand{\tev}{\aunit{Te\kern -0.1em V}\xspace}
\newcommand{\gev}{\aunit{Ge\kern -0.1em V}\xspace}
\newcommand{\mev}{\aunit{Me\kern -0.1em V}\xspace}
\newcommand{\kev}{\aunit{ke\kern -0.1em V}\xspace}
\newcommand{\ev}{\aunit{e\kern -0.1em V}\xspace}
 
\newcommand{\mevc}{\ensuremath{\aunit{Me\kern -0.1em V\!/}c}\xspace}
\newcommand{\gevc}{\ensuremath{\aunit{Ge\kern -0.1em V\!/}c}\xspace}
\newcommand{\mevcc}{\ensuremath{\aunit{Me\kern -0.1em V\!/}c^2}\xspace}
\newcommand{\gevcc}{\ensuremath{\aunit{Ge\kern -0.1em V\!/}c^2}\xspace}

\newcommand{\chisqndf}{\ensuremath{\chi^2/\mathrm{ndf}}\xspace}

\def\gsim{{~\raise.15em\hbox{$>$}\kern-.85em
          \lower.35em\hbox{$\sim$}~}\xspace}
\def\lsim{{~\raise.15em\hbox{$<$}\kern-.85em
          \lower.35em\hbox{$\sim$}~}\xspace}

\def\tell1  {TELL1\xspace}
\def\ukl1   {UKL1\xspace}

\def\toa{\mbox{\textrm{ToA}}\xspace} 
\def\tot{\mbox{\textrm{ToT}}\xspace}

\def\timepixfour{Timepix4\xspace}

\DeclareSIUnit{\clight}{\mathit{c}}

\title{Neural network-based timewalk correction for the Timepix4 ASIC}

\author[a,1]{D.~Bacher\note{Corresponding author.}} 
\author[b]{K.~Akiba} 
\author[b]{M.~van~Beuzekom} 
\author[c]{V.~Coco}  
\author[c]{R.~Dumps} 
\author[b]{T.~Evans} 
\author[b]{K.~Heijhoff} 
\author[a]{M.~John} 
\author[c]{E.~Lemos~Cid} 
\author[c]{T.~Pajero} 

\affiliation[a]{{Department of Physics, University of Oxford, Denys Wilkinson Bldg., Keble Road, Oxford, OX1~3RH, United Kingdom}}
\affiliation[b]{{Nikhef, Science Park 105, 1098 XG Amsterdam, the Netherlands}}
\affiliation[c]{{CERN, Esplanade des Particules 1, 1211 Geneva, Switzerland}}

\emailAdd{david.bacher@physics.ox.ac.uk}

\abstract{Timewalk corrections of hybrid pixel detectors conventionally employ heuristic fits performed independently in bins of the relevant input variables.
This requires both an explicit functional ansatz and binning scheme, as well as densely populated calibration data whose required size grows rapidly as further input variables are added.
This paper presents a timewalk correction for the \timepixfour readout chip based on a compact feed-forward neural network (NN), which learns the dependence directly from the input variables without requiring an explicit functional ansatz or binning.
The NN is compared against the fit-based baseline using data from the \timepixfour beam telescope at the \cern SPS in 2025, with \qty{100}{\micro\meter} and \qty{300}{\micro\meter} thick planar sensors.
Both methods parameterise the timewalk correction as a function of the measured charge and reconstructed intrapixel impact position.
With only \qty{20}{\percent} of the calibration data, the NN already achieves the same asymptotic time resolutions as the fit-based baseline: \qty{180}{\pico\second} and \qty{300}{\pico\second} for the two thicknesses, respectively.
}


\graphicspath{{./figs/}}

\begin{document}
\maketitle
\flushbottom


\section{Introduction}\label{sec:introduction}
Hybrid pixel detectors, in which a finely segmented silicon sensor is bump-bonded to a dedicated readout chip, are the primary technology for charged-particle tracking and vertexing at collider experiments such as at the LHC~\cite{Garcia-Sciveres:2017ymt}.
Adding picosecond-level per-hit timing to such detectors is a major area of current research since it is required to enable four-dimensional (4D) tracking, in which timing information helps resolve the many overlapping collisions at high-luminosity colliders~\cite{LHCbcollaboration:2903094}.
Reaching this timing performance requires calibrating each pixel to remove several instrumental effects~\cite{Akiba:2025flo}, a regression problem in which the recorded time depends, in a complex and non-linear way, on the deposited charge and other operating and external conditions.
The conventional solution is a heuristic fit performed independently in many bins, repeated for every pixel, which requires large, densely populated data sets and fails where data are scarce or unevenly distributed.
Machine-learning methods have been applied extensively to reconstruction problems in high-energy physics~\cite{Albertsson:2018maf}, including cluster reconstruction in silicon pixel detectors~\cite{ATLAS:2014lvy}.
These applications motivate investigating whether neural networks (NNs) can also be applied for the timing calibrations of pixel detectors.
This paper studies how a compact NN can learn such a timing calibration directly from data, and how its accuracy depends on the volume of training data and the complexity of the model.

The timing performance of silicon pixel sensors and its dependence on the sensor geometry have been studied in detail in Ref.~\cite{Riegler:2017xbh}.
The dominant instrumental effect is timewalk, a charge-dependent bias common to detectors with a fixed-threshold discriminator: since the signal rise time is approximately constant, a larger charge crosses the threshold earlier, biasing the measured Time-of-Arrival (\toa) for signals produced at the same time~\cite{Riegler:2017xbh}.
A further effect arises from the sensor's non-uniform weighting field, which makes the centroid time of a signal depend on where within the pixel the deposited charge is distributed~\cite{Riegler:2017xbh}.
In a planar sensor, the effect mainly depends on the depth of the charge deposition which depends on the impact point and incidence angle of the charged particle.
In beam telescopes, the incidence angle is fixed with respect to the telescope planes and thus the timewalk correction can be parameterised as a function of the measured charge and the reconstructed intrapixel impact position~\cite{Heijhoff:2020mlk}.
The conventional approach fits this dependence with a heuristic function, independently in many bins of the impact position.

In this paper, the heuristic fit is replaced by a compact feed-forward NN that takes the measured charge and the intrapixel impact position as inputs and predicts the timewalk correction directly.
It shares its inputs with the fit-based method, so the two approaches can be compared directly.
The central methodological advantage is that the NN learns the multidimensional dependence from the data without requiring either an explicit functional ansatz or a choice of intrapixel binning.
Additionally, further observables and their correlations can easily be included as additional inputs without constructing increasingly high-dimensional bins.
This study deliberately uses the same inputs for the NN and the fit-based method, allowing their performance and data requirements to be compared when both methods have access to the same information.

The remainder of the paper is organised as follows.
After a description of the \timepixfour beam telescope and data set in \cref{sec:tpx4_telescope}, the fit-based baseline and the dependence of the timewalk on the charge deposition position are presented in \cref{sec:setup}.
The NN correction and its training are introduced in \cref{sec:nn_timewalk_correction}.
Its dependence on model complexity and calibration data volume, together with a comparison against the fit-based method is reported in \cref{sec:results}.
The results and possible extensions of the method are discussed in \cref{sec:conclusion}.

\section{The Timepix4 beam telescope}\label{sec:tpx4_telescope}
The two methods are compared using data recorded with the \timepixfour beam telescope, described in detail in Ref.~\cite{Akiba:2025flo}, at the \cern SPS H8 beamline~\cite{sps-h8} with a \qty{180}{\giga\electronvolt\per\clight} mixed hadron beam.
The telescope is based on the \timepixfour application specific integrated circuit (ASIC)~\cite{llopart2022timepix4} bump bonded to planar silicon sensors.
The ASIC has a matrix of \numproduct{448 x 512} pixels, each \qtyproduct{55x55}{\micro\meter} in size, giving \num{229376} pixels in total. Each pixel provides a \toa measurement with a nominal binning of \qty{195}{\pico\second} and a Time-over-Threshold (\tot) measurement with \qty{1.56}{\nano\second} binning.
The timing performance of the bare \timepixfour ASIC and bump-bonded assemblies has been extensively characterised in Refs.~\cite{Heijhoff_2022,Bolzonella:2024}.

The telescope comprises eight tracking planes, four with \qty{100}{\micro\meter} thick planar sensors and four with \qty{300}{\micro\meter} thick planar sensors.
The \qty{100}{\micro\meter} planes are oriented at perpendicular incidence, while the \qty{300}{\micro\meter} planes are tilted by \qty{9}{\degree} about both $x$ and $y$ to improve the spatial resolution\footnote{The coordinate system follows the definition in Ref.~\cite{Akiba:2025flo} with the $z$-axis along the beam direction and the $y$-axis pointing upwards.}.
The telescope is used to reconstruct the tracks of charged particles crossing the sensors.
The intrapixel impact position is computed as the intersection of the reconstructed tracks with individual planes and is used to study the timewalk behaviour and to train the NN.
The telescope achieves a pointing resolution of \qty{2.3\pm0.1}{\micro\meter} at its centre and about \qty{6}{\micro\meter} at the outer planes.
The time reference is provided by two micro-channel plate photomultiplier tubes with a resolution of approximately \qty{12}{\pico\second}.

The tilting of the \qty{300}{\micro\meter} planes causes the depth at which a particle traverses a given pixel to depend on its position within that pixel.
Consequently, the timewalk correction is expected to depend on the intrapixel position, defined relative to the centre of the pixel under consideration~\cite{Heijhoff:2020mlk}.
For the \qty{100}{\micro\meter} planes at perpendicular incidence, all particles cross the sensor in the same way, so the timewalk correction is expected to be largely independent of the intrapixel position.

Data were collected in July 2025 and reconstructed using the \textsc{Kepler}\footnote{\url{https://gitlab.cern.ch/TimepixTelescope/Kepler/}} software.
The data set holds more than \num{5000} hits per pixel in a \numproduct{40 x 60} pixel region on each plane, allowing performance to be evaluated as a function of the number of calibration hits used per pixel.
Before any timewalk correction is applied, frequency variations of the voltage controlled oscillators (VCOs) in the \timepixfour{} ASICs are accounted for, as described in Ref.~\cite{Akiba:2025flo}.
The timestamp of a reconstructed cluster is given by the \toa of the earliest pixel in the cluster, which is defined as the leading pixel.
For the corrections presented in the following sections, the intrapixel position is defined relative to the centre of the leading pixel, and the charge is taken from the leading pixel only.

\section{Fit-based timewalk correction}\label{sec:setup}
The conventional timewalk correction fits a heuristic function to the measured \toa as a function of the collected charge.
The charge is obtained from the \tot measurement using a per-pixel calibration derived from test pulse data as described in Ref.~\cite{Akiba:2025flo}.
To capture variations in signal collection across the pixel, the fit is performed in bins of the reconstructed track intrapixel position in $x$ and $y$, with a bin size of \qtyproduct{2 x 2}{\micro\meter} chosen to match the pointing resolution of the telescope.

The correction proceeds in three steps:
\begin{enumerate}
    \item A per-pixel offset correction equalises the \toa distribution across the pixel matrix.
    \item The measured \toa as a function of charge is fitted in bins of the intrapixel impact position.
    \item A second per-pixel offset correction removes any remaining pixel-to-pixel variations.
\end{enumerate}
The correction function used in step~2 has the form
\begin{equation}
    \Delta t_{\text{timewalk}}(q) = a + \frac{b}{q^c}\,,
\end{equation}
where $q$ is the measured charge and $a$, $b$, and $c$ are determined independently in each bin.
Fits with $\chisqndf < 20$ are considered successful.
Bins that fail to converge or do not satisfy this criterion are excluded from the analysis.
With sufficiently large data samples this happens in less than \qty{1}{\percent} of the bins.

\subsection{Dependence on intrapixel position}\label{sec:setup:intrapixel}
\begin{figure}
    \centering
    \begin{subfigure}{0.5\textwidth}
        \centering
        \includegraphics[width=0.9\textwidth]{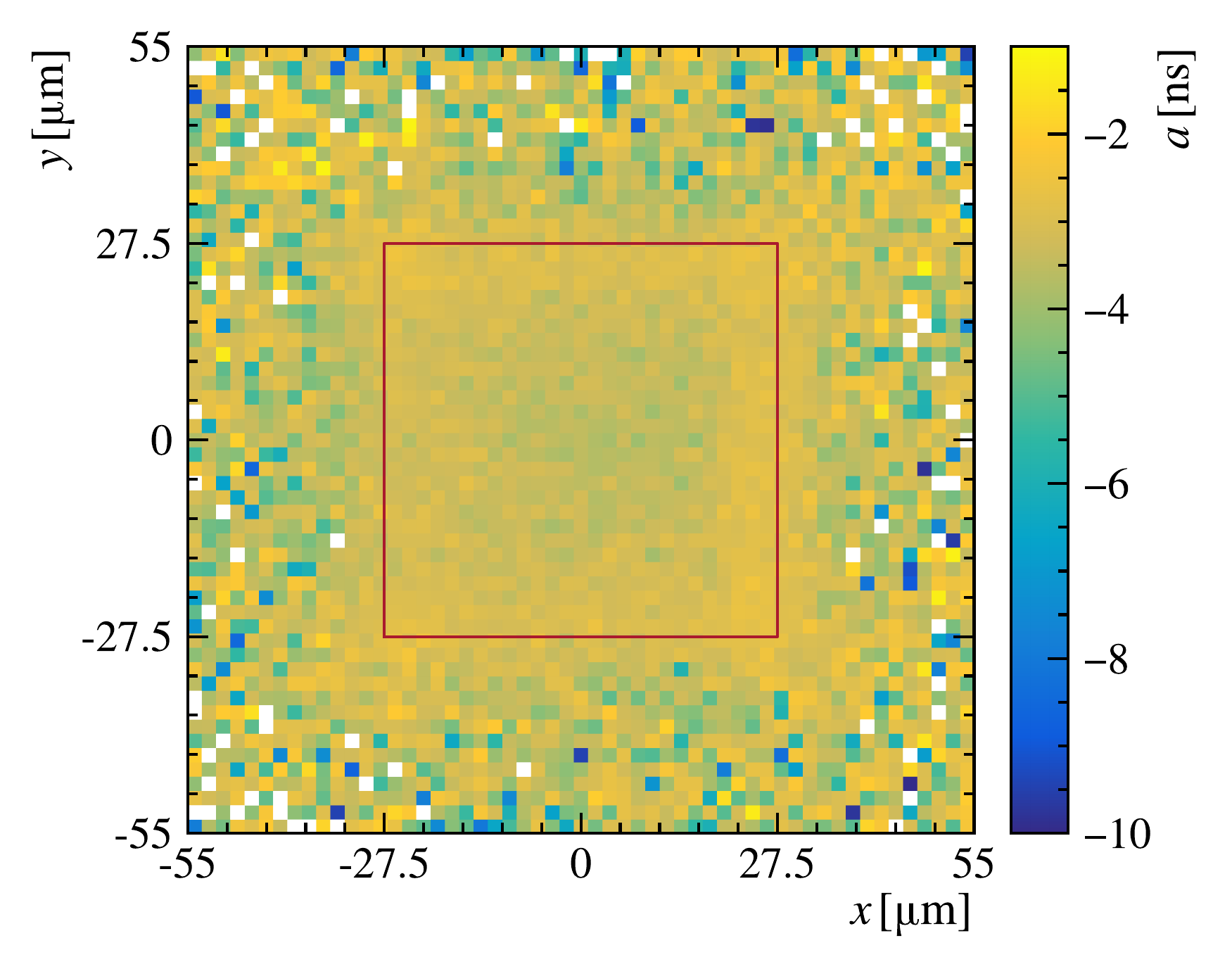}
        \caption{Parameter $a$}\label{fig:timewalk_correction_bins_100:a}
    \end{subfigure}%
    \hfill%
    \begin{subfigure}{0.5\textwidth}
        \centering
        \includegraphics[width=0.9\textwidth]{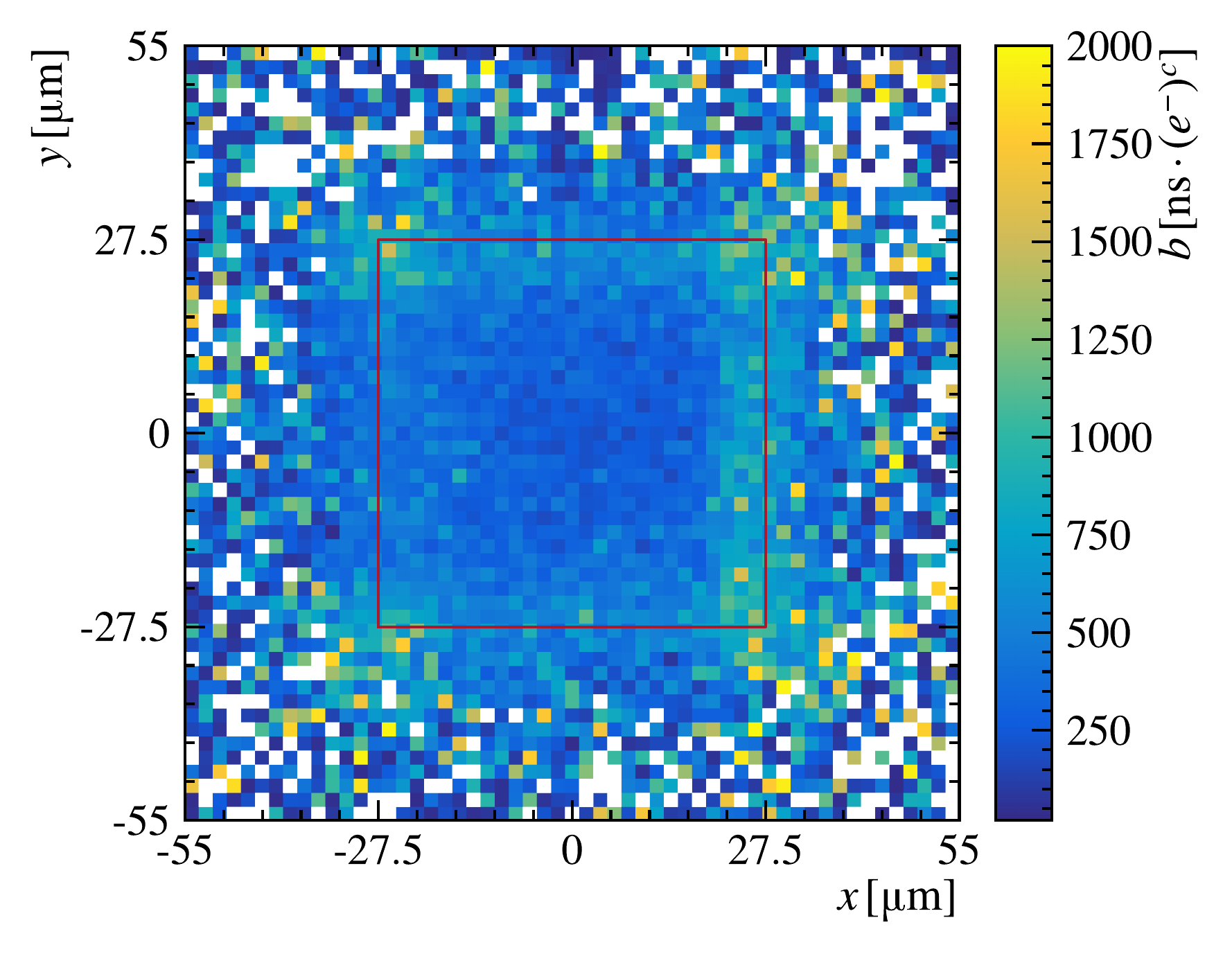}
        \caption{Parameter $b$}\label{fig:timewalk_correction_bins_100:b}
    \end{subfigure}
    \begin{subfigure}{0.5\textwidth}
        \centering
        \includegraphics[width=0.9\textwidth]{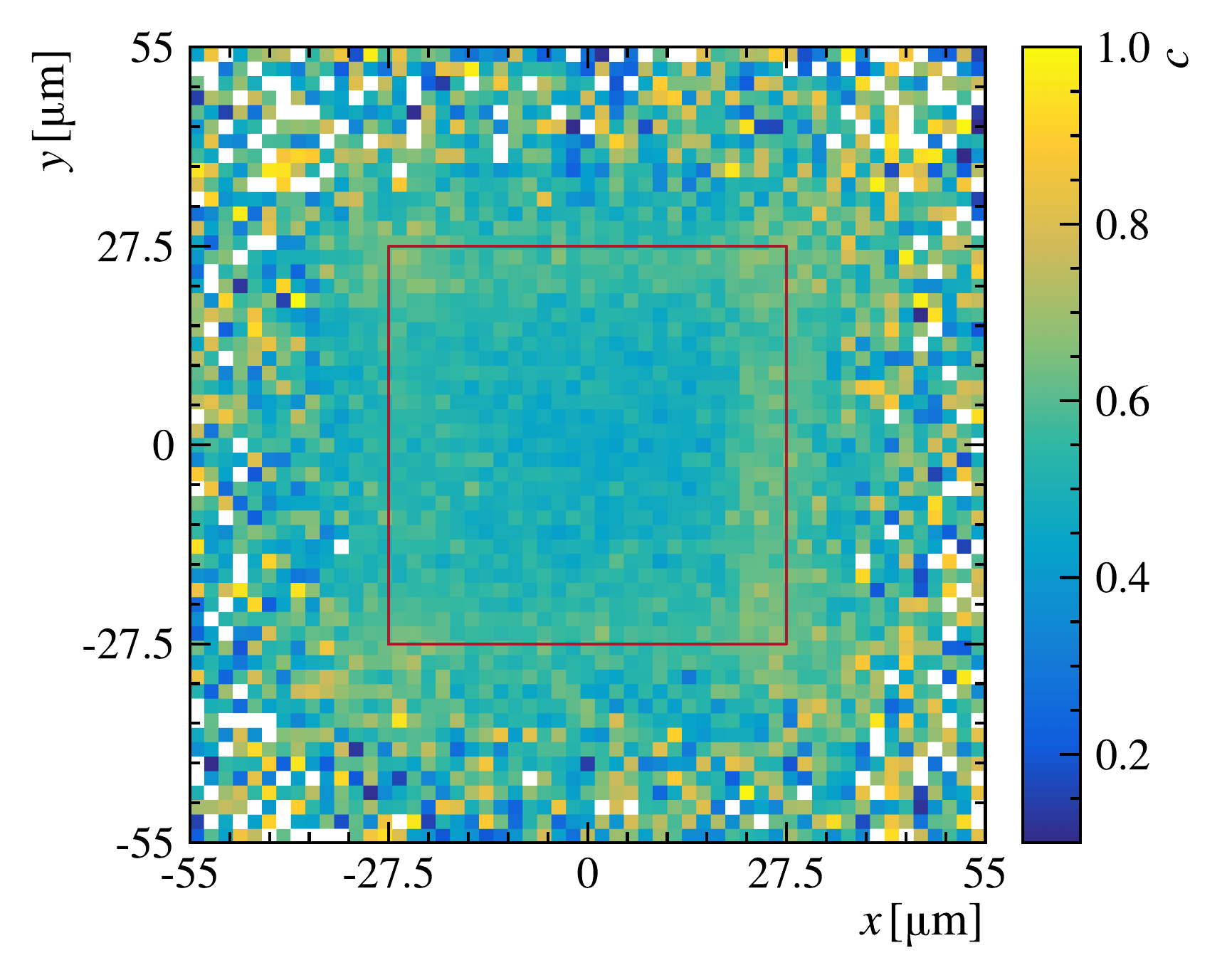}
        \caption{Parameter $c$}\label{fig:timewalk_correction_bins_100:c}
    \end{subfigure}%
    \hfill%
    \begin{subfigure}{0.5\textwidth}
        \centering
        \includegraphics[width=0.9\textwidth]{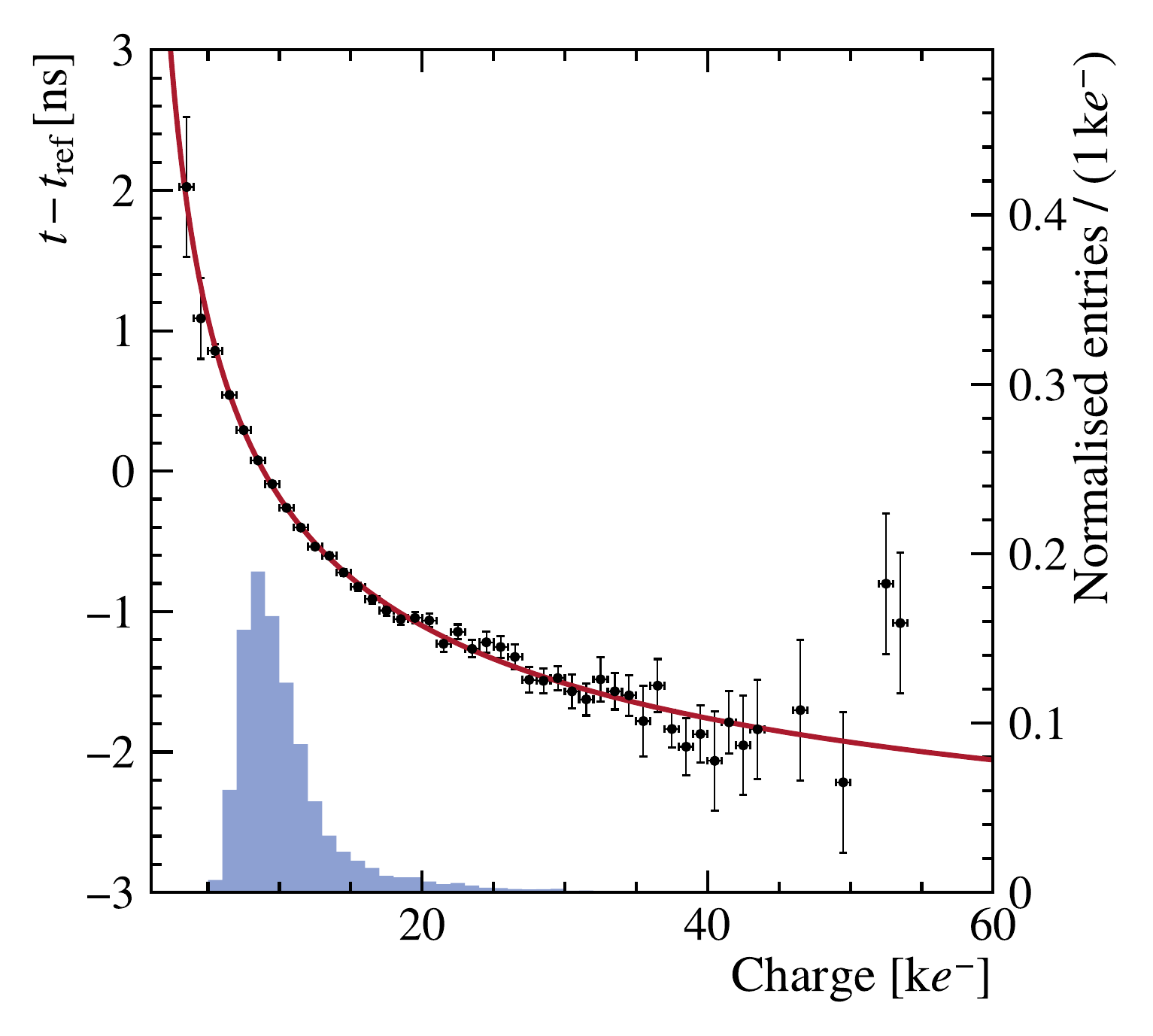}
        \caption{Example fit in the central bin}\label{fig:timewalk_correction_bins_100:example}
    \end{subfigure}
    \caption{\subref{fig:timewalk_correction_bins_100:a}--\subref{fig:timewalk_correction_bins_100:c}: Fitted parameters $a$, $b$, and $c$ of the timewalk correction function as a function of the intrapixel impact position for a \qty{100}{\micro\meter} plane at perpendicular incidence.
    The pixel region is indicated by the red square.
    \subref{fig:timewalk_correction_bins_100:example}: Example fit in the central bin of the intrapixel position.
    The distribution of the measured charge is shown by the blue histogram on the right-hand axis.
    }\label{fig:timewalk_correction_bins_100}
\end{figure}

The fitted parameters $a$, $b$, and $c$ across the intrapixel position bins are shown together with an example fit in the central bin in \cref{fig:timewalk_correction_bins_100} for a \qty{100}{\micro\meter} plane.
No significant intrapixel dependence is observed, as expected for perpendicular incidence.
A slight change in the parameters $b$ and $c$ is observed close to the pixel edges, where the weighting field is less homogenous.

\begin{figure}
    \centering
    \begin{subfigure}{0.5\textwidth}
        \centering
        \includegraphics[width=0.9\textwidth]{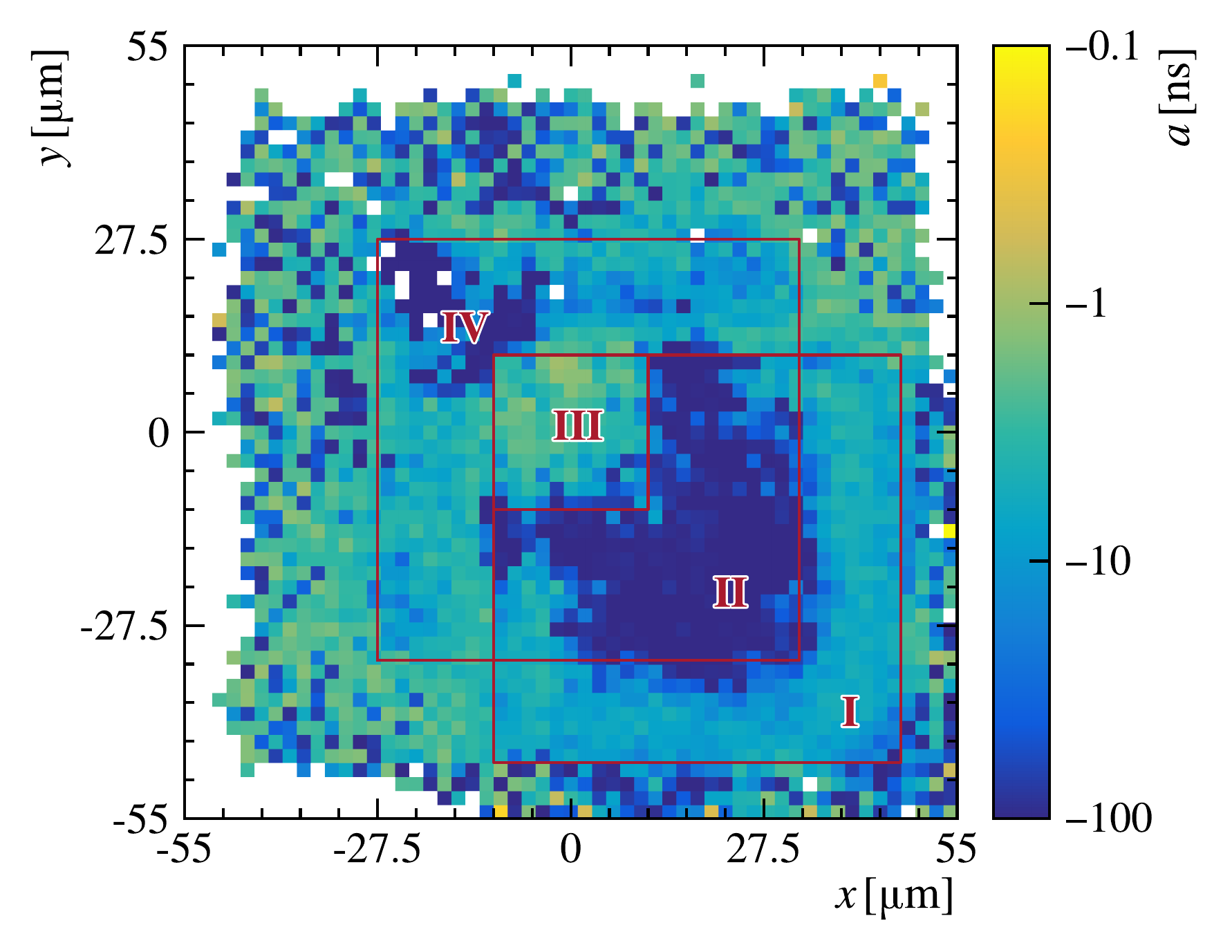}
        \caption{Parameter $a$}\label{fig:timewalk_correction_bins_300:a}
    \end{subfigure}%
    \hfill%
    \begin{subfigure}{0.5\textwidth}
        \centering
        \includegraphics[width=0.9\textwidth]{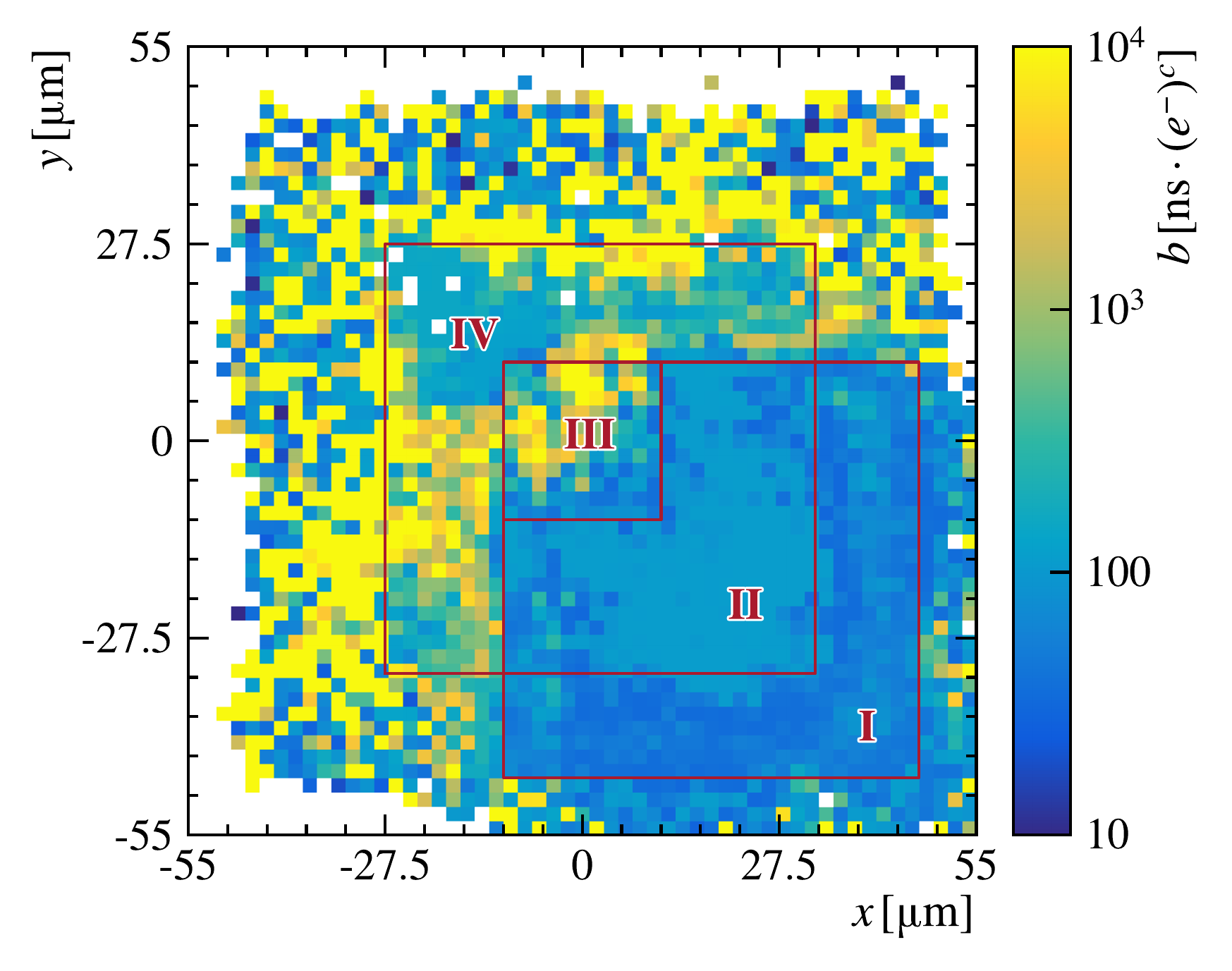}
        \caption{Parameter $b$}\label{fig:timewalk_correction_bins_300:b}
    \end{subfigure}
    \begin{subfigure}{0.5\textwidth}
        \centering
        \includegraphics[width=0.9\textwidth]{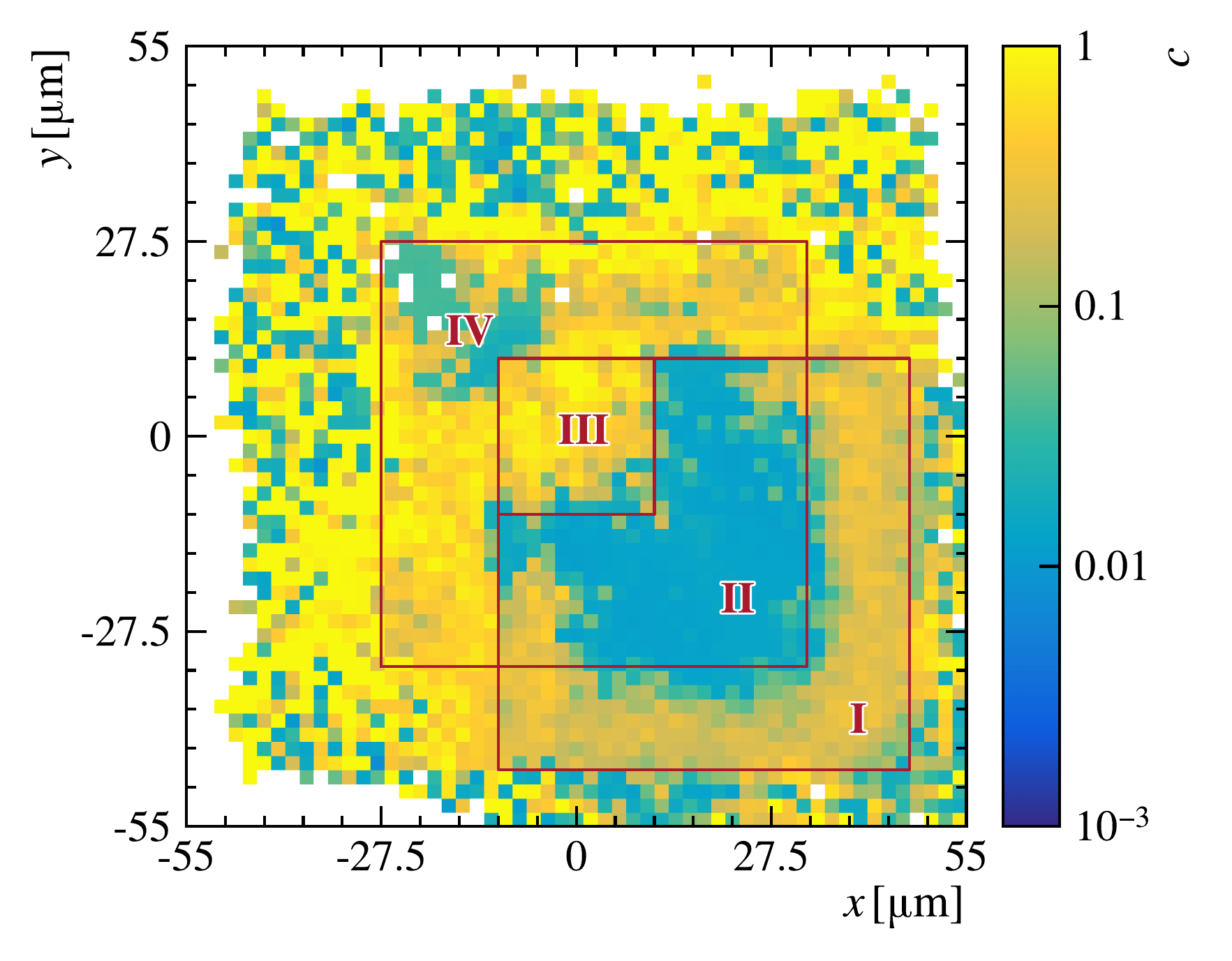}
        \caption{Parameter $c$}\label{fig:timewalk_correction_bins_300:c}
    \end{subfigure}
    \caption{Fitted parameters $a$, $b$, and $c$ of the timewalk correction function as a function of the intrapixel impact position for a \qty{300}{\micro\meter} plane tilted by \qty{9}{\degree} around both the $x$ and $y$ axes.
    }\label{fig:timewalk_correction_bins_300}
\end{figure}

For the \qty{300}{\micro\meter} planes, the timewalk behaviour is expected to depend on the intrapixel position, since the particles cross the pixel at different depths depending on the impact position.
The fitted parameters $a$, $b$, and $c$ are shown in \cref{fig:timewalk_correction_bins_300}.
Because the plane is tilted, the shadow of the pixel is shifted towards the bottom right: the intrapixel position denotes the point at which the particle crosses the mid-plane of the sensor, which is offset from the readout electrode.

\begin{figure}
    \centering
    \begin{subfigure}{0.5\textwidth}
        \centering
        \includegraphics[width=0.9\textwidth]{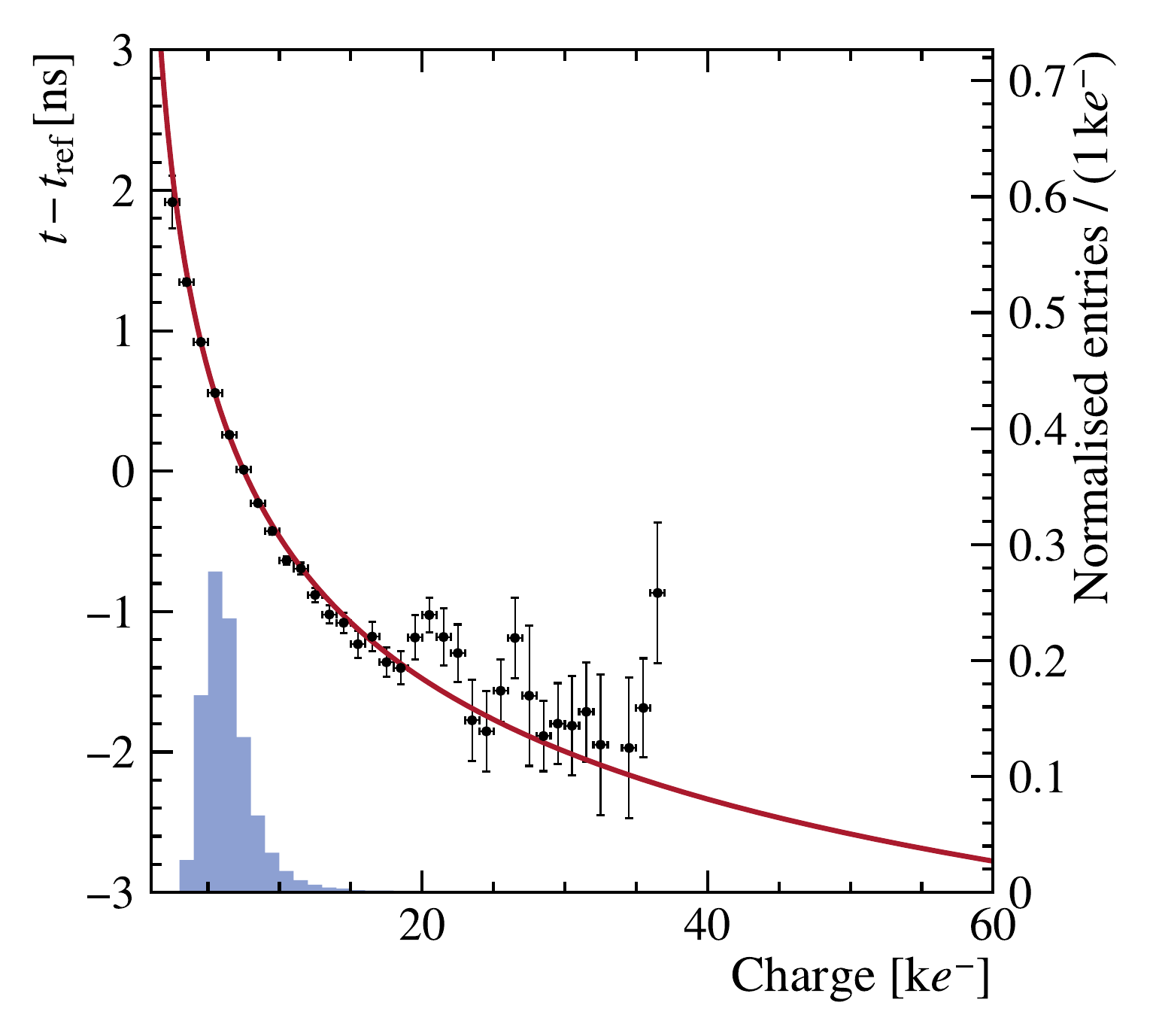}
        \caption{Region I}\label{fig:timewalk_correction_bins_300_example_fits:I}
    \end{subfigure}%
    \hfill%
    \begin{subfigure}{0.5\textwidth}
        \centering
        \includegraphics[width=0.9\textwidth]{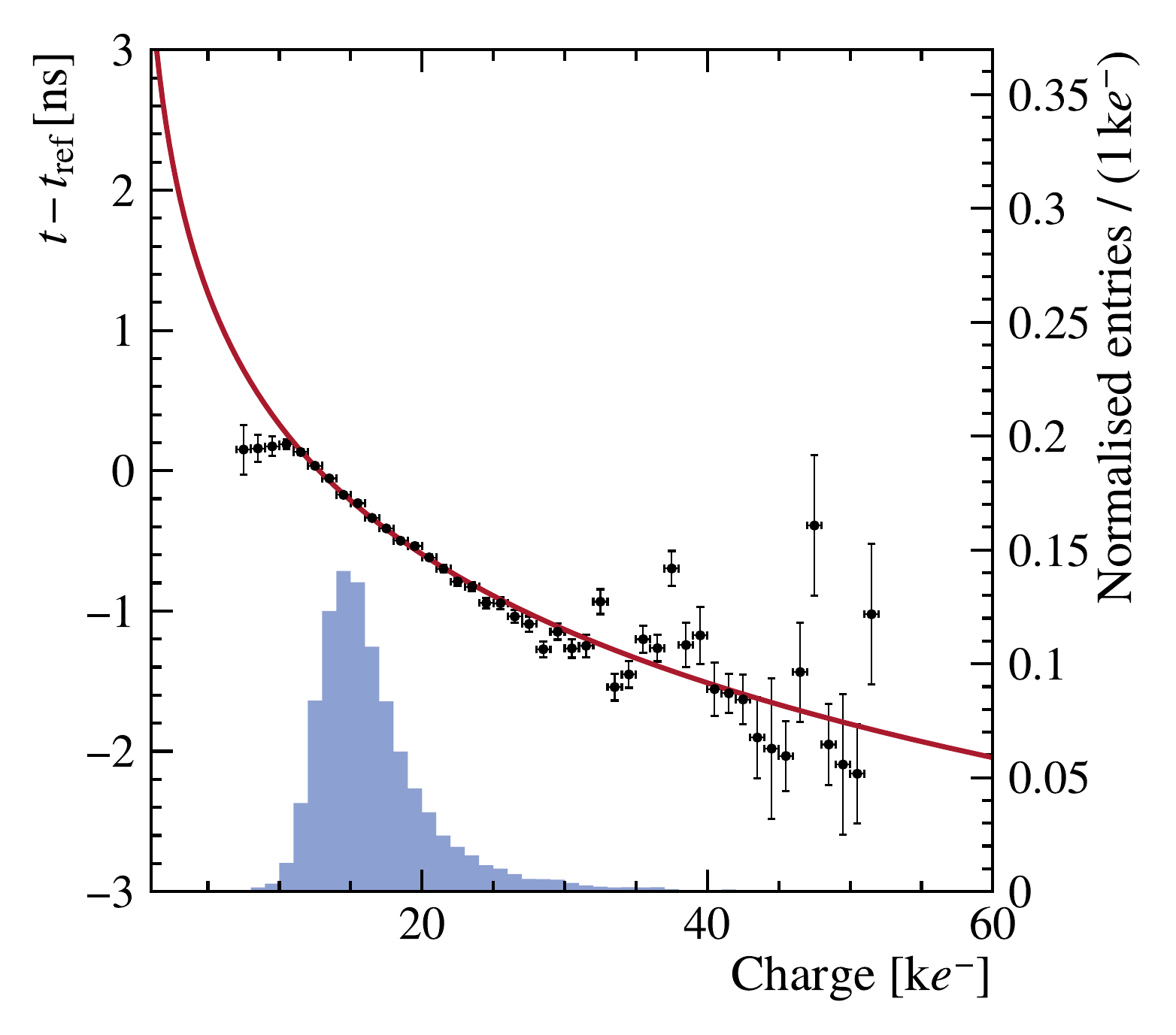}
        \caption{Region II}\label{fig:timewalk_correction_bins_300_example_fits:II}
    \end{subfigure}
    \begin{subfigure}{0.5\textwidth}
        \centering
        \includegraphics[width=0.9\textwidth]{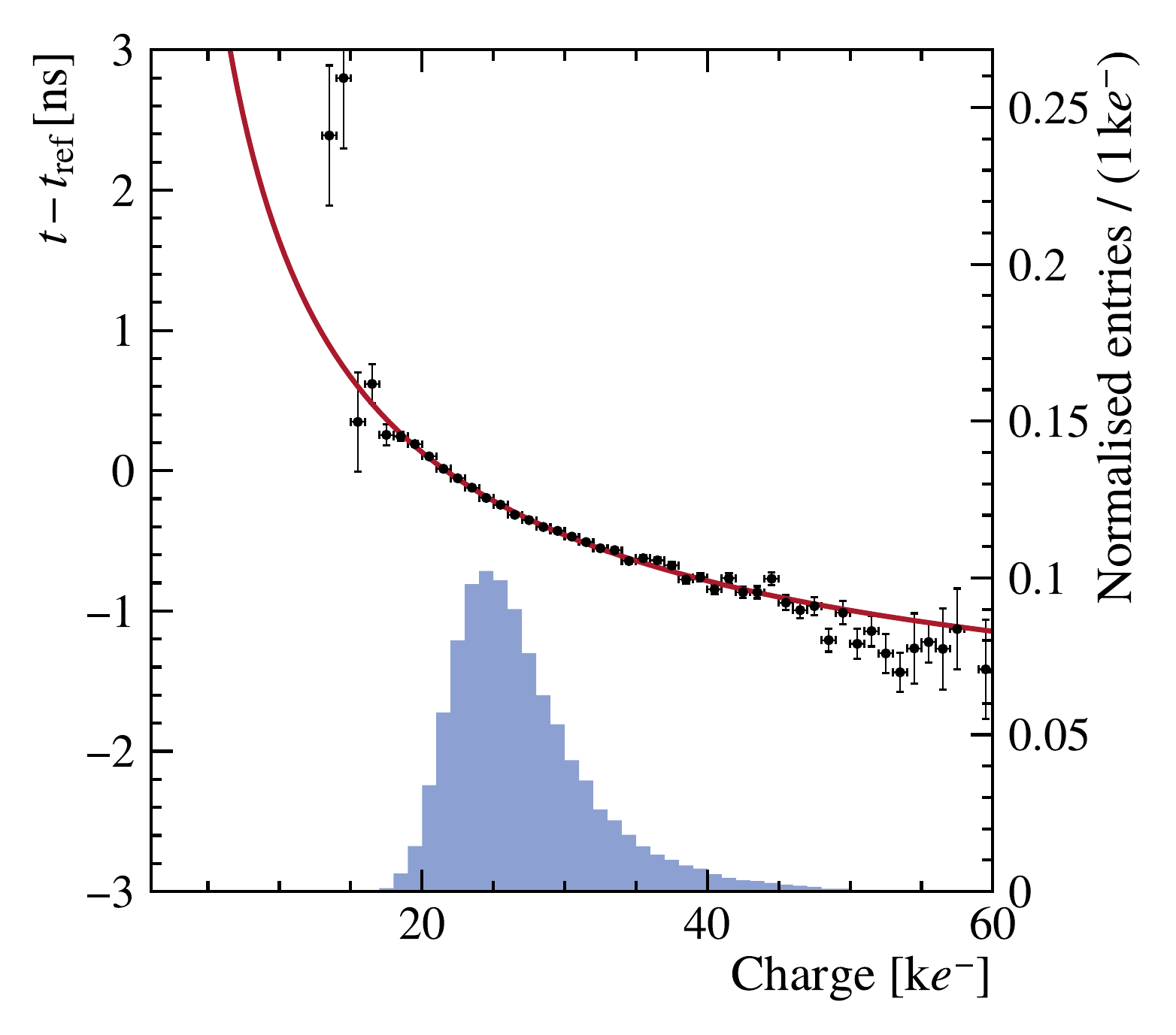}
        \caption{Region III}\label{fig:timewalk_correction_bins_300_example_fits:III}
    \end{subfigure}
    \caption{Example fits in the regions I--III of the intrapixel position, as defined in \cref{fig:timewalk_correction_bins_300}, for a \qty{300}{\micro\meter} plane.
    The distribution of the measured charge is shown by the blue histogram on the right-hand axis.
    }\label{fig:timewalk_correction_bins_300_example_fits}
\end{figure}

\begin{figure}
    \centering
    \includegraphics[width=0.3\textwidth]{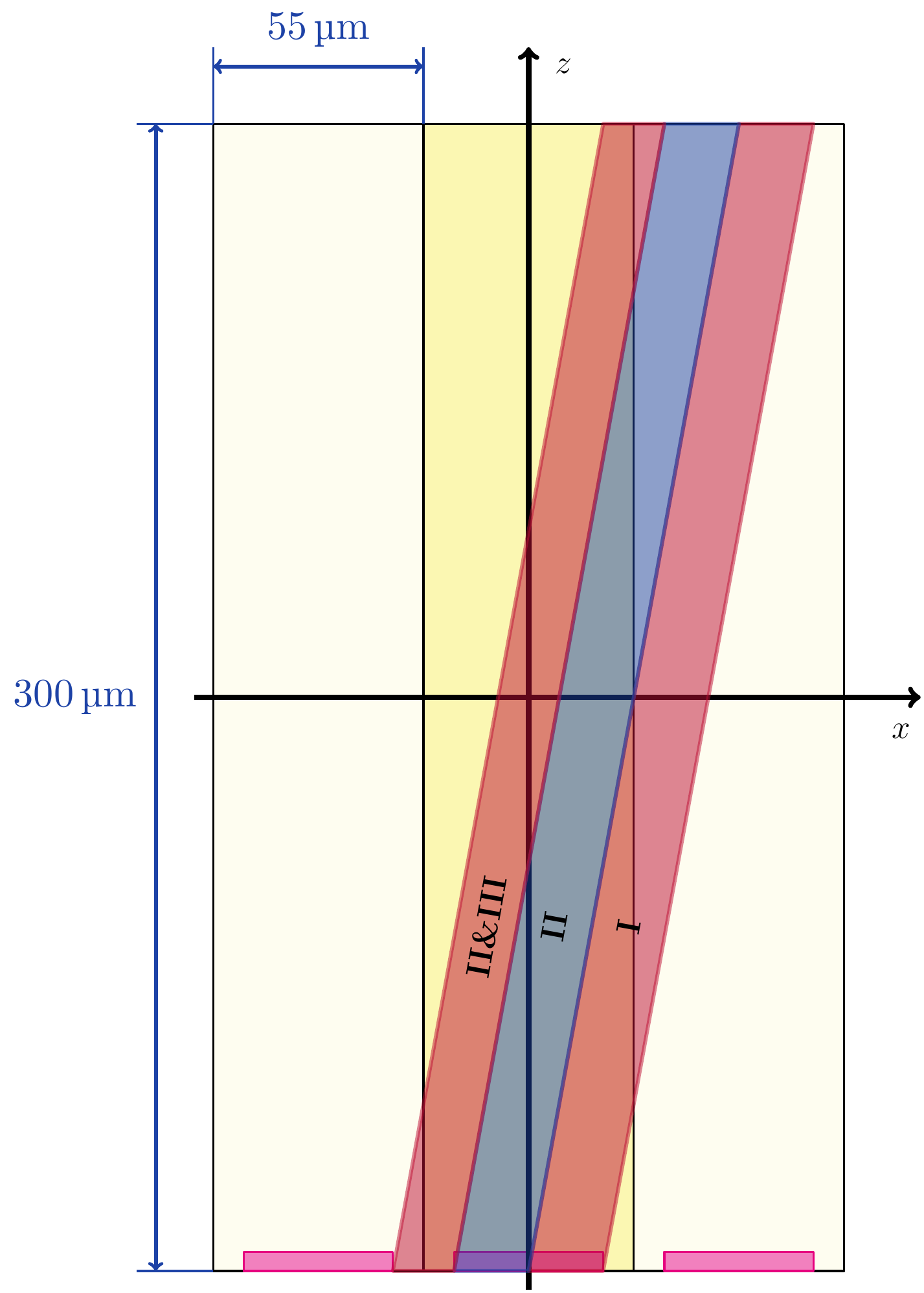}
    \caption{Illustration of a \qty{300}{\micro\meter} plane, showing the leading pixel (centre) and its neighbouring pixels to the left and right, with the readout electrodes indicated by the pink rectangles.
    The different regions of the intrapixel position, labelled I--III as indicated in \cref{fig:timewalk_correction_bins_300}, correspond to different charge-sharing scenarios between the leading pixel and its neighbours, as well as to different distances from the readout electrode.
    The intrapixel position corresponds to the particle's position at the mid-point of the sensor depth.
    }\label{fig:regions}
\end{figure}

As previously found in Ref.~\cite{Heijhoff:2020mlk}, three main regions are identified, labelled I--III in \cref{fig:timewalk_correction_bins_300}, for each of which an example fit is shown in \cref{fig:timewalk_correction_bins_300_example_fits}.
As illustrated in \cref{fig:regions}, these regions correspond to different charge-sharing scenarios between the leading pixel and its neighbours, and to different distances of the particle crossing point from the readout electrode.
Region~I corresponds to a small charge deposit in the leading pixel, close to its readout electrode, with most of the charge deposited in neighbouring pixels.
Region~II corresponds to the majority of the charge being deposited in the leading pixel, with the particle crossing close to its readout electrode.
Region~III corresponds to almost all the charge being deposited in the leading pixel, but with the particle crossing in the corner of the pixel, further from the readout electrode.
A fourth region, IV, is also observed, corresponding to the opposite case of region~I, where the pixel receiving the largest charge deposit is the leading pixel.
This scenario rarely occurs ($< \qty{1}{\percent}$ of hits) and thus has a negligible impact.
The points outside regions I--IV occur even less and are statistical outliers without any defined or common behaviour.

\section{Neural-network timewalk correction}\label{sec:nn_timewalk_correction}
The fit-based procedure requires both a functional ansatz and sufficient calibration hits to constrain it in every intrapixel bin.
The NN-based procedure avoids the functional ansatz and removes the need to populate every intrapixel bin by learning a continuous multidimensional correction directly from the calibration data.
It follows the same three-step structure as the baseline, but replaces the bin-by-bin fit in step~2 with a NN that predicts the correction directly.
A separate NN is trained for each detector plane, while all pixels within a plane share the same network.

\subsection{Network and inputs}
The NN is a fully connected, feed-forward multi-layer perceptron.
Its inputs are the measured charge~$q$ and the reconstructed intrapixel impact position in $x$ and $y$ relative to the pixel centre.
These are deliberately the same inputs used by the fit-based correction so that the comparison isolates the correction method and its data requirement.
The inclusion of further variables and their potential correlations is not investigated in this study, but can be straightforwardly implemented.
Its single output is the predicted correction $\Delta t_{\text{NN}}$, which is subtracted from the measured \toa.
The hidden layers use a rectified-linear (ReLU) activation function, and the output layer is linear.
The depth (number of hidden layers) and width (neurons per hidden layer) are free hyperparameters whose influence is studied in \cref{sec:results}.
Unless stated otherwise, the results use two hidden layers of sixteen neurons, identified in \cref{sec:results} as a robust compromise between achieved performance and training stability.

\subsection{Training}
\begin{figure}
    \centering
    \includegraphics[width=0.45\textwidth]{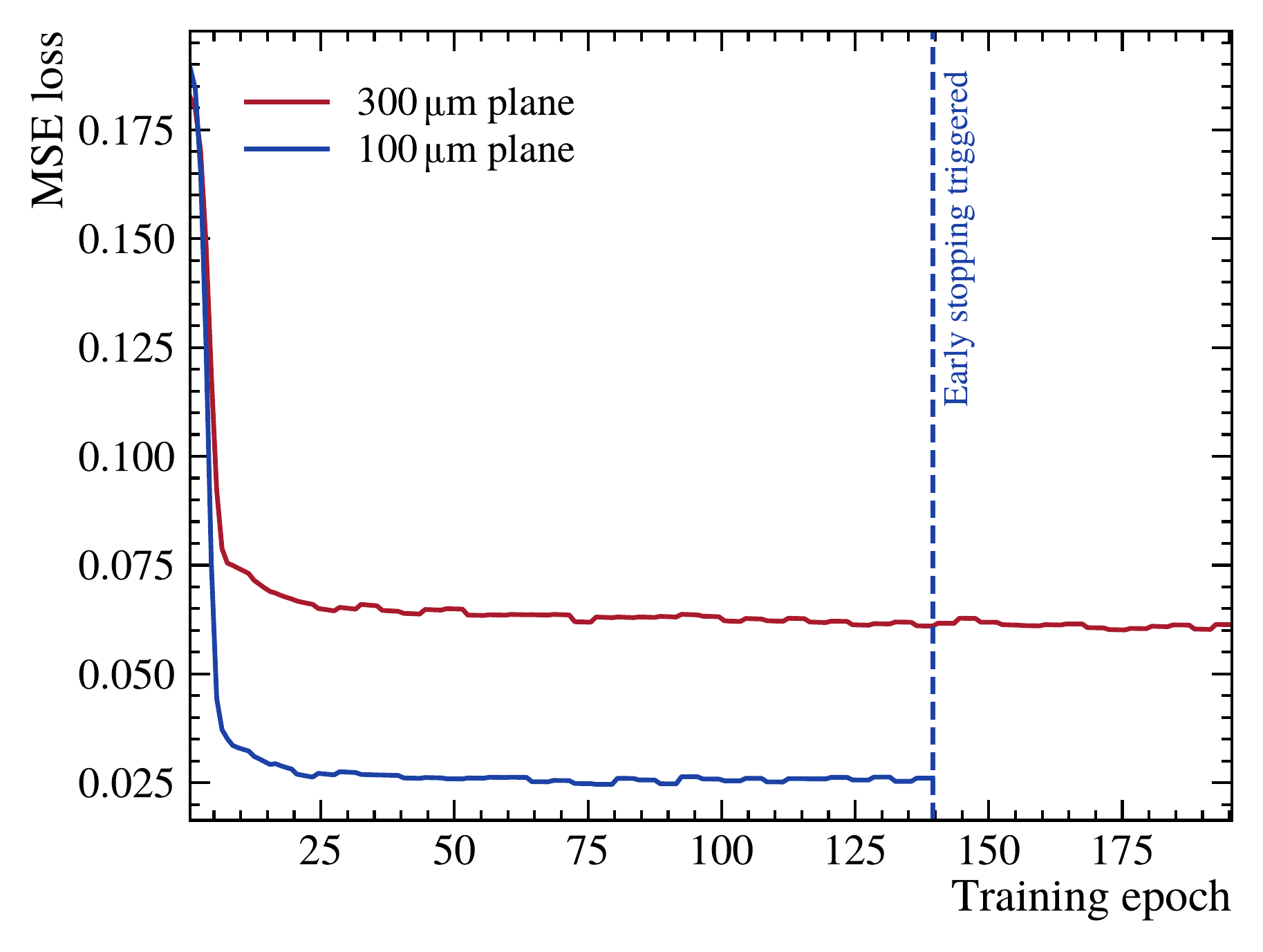}
    \caption{Validation loss as a function of the number of epochs processed (integrated over successive batches) for the NN-based timewalk correction for a \qty{100}{\micro\meter} (blue) and \qty{300}{\micro\meter} (red) plane.
    The discrete fluctuations correspond to the individual training batches, since a new validation set is used in each batch.
    }\label{fig:training_loss}
\end{figure}

Training uses the Adam optimiser~\cite{Adam}, a mean squared error (MSE) loss, and a learning rate of \num{e-3}.
Hits whose time residuals exceed ten times the median absolute deviation are rejected as outliers to avoid biasing the training.
The data are processed in successive batches of consecutive hits, with each batch trained for several epochs before moving on to the next.
This batch-wise processing naturally supports an online scenario, where calibration hits arrive continuously and the model is updated as new data are collected, although this is not investigated in this study.
Training stops when the loss on a held-out validation set fails to improve for a set number of epochs, or when a maximum number of batches is reached.
In each batch the training is performed on \qty{90}{\percent} of the data, while the remaining \qty{10}{\percent} form the validation set.
The performance of the trained NN is evaluated on a separate test data sample, which is not used for training or validation.
An example of the training loss is shown in \cref{fig:training_loss}.

\section{Results}\label{sec:results}
The NN-based correction is evaluated in two ways.
First, its dependence on model complexity is mapped with a scan over the NN width and depth, in order to identify a suitable default architecture (\cref{sec:results:complexity}).
Second, the performance and efficiency of this architecture are compared against the fit-based baseline as a function of the number of calibration hits per pixel (\cref{sec:results:efficiency}).

Two metrics are used throughout: The first is the time resolution, defined as the root-mean-square~(RMS) of the central 99th percentile of the time differences between the corrected \toa and the reference time, with the known \qty{12}{\pico\second} contribution from the time reference system subtracted in quadrature.
The second is the correction efficiency, the fraction of clusters for which a correction can be obtained.
For the fit-based correction, the bin-by-bin fit can fail to converge in sparsely populated bins, leaving those hits uncorrected.
For this method, the time resolution is evaluated using only clusters for which the bin-by-bin fit provides a valid correction.
The reported time resolution and correction efficiency must therefore be considered together, particularly at low calibration sample sizes.
The NN, by contrast, is defined everywhere and returns a correction for every hit, so its correction efficiency is \qty{100}{\percent} in all cases, independent of the network architecture.
When comparing the time resolution of the two methods, only clusters for which the fit-based correction is defined are used.
No degradation of the time resolution from the NN is observed when including all clusters.

Both correction methods use the same per-pixel offset correction in step~1, while the second offset correction in step~3 is then computed independently for each method, to absorb any residual pixel-to-pixel differences left by its respective timewalk correction.
Both per-pixel offset corrections are derived using the same number of calibration hits as the timewalk correction at each point in the scan.
The two correction methods therefore are based on calibration sample of equal size throughout.

The same qualitative trends are observed for the different planes of each thickness.
Results are therefore shown for one representative plane of each thickness.

\subsection{Dependence on model complexity}\label{sec:results:complexity}
The accuracy of a learned correction depends on the capacity of the NN: too few parameters and it cannot capture the timewalk dependence, too many, and it may overfit or fail to converge when data are scarce.
To map this trade-off, a NN is trained for every combination\footnote{All hidden layers have equal size.} of width (neurons per layer) $\in\{4,8,16,32,64\}$ and depth (hidden layers) $\in\{1,2,3\}$, using the same training configuration throughout, and its time resolution is evaluated for each combination.
Architectures are denoted $d$L$w$N, where $d$ is the depth and $w$ the width, so that, for example, 2L16N refers to a NN with two hidden layers of sixteen neurons each.

\begin{figure}
    \centering
    \begin{subfigure}{\textwidth}
        \begin{subfigure}{0.5\textwidth}
            \centering
            \includegraphics[width=0.9\textwidth]{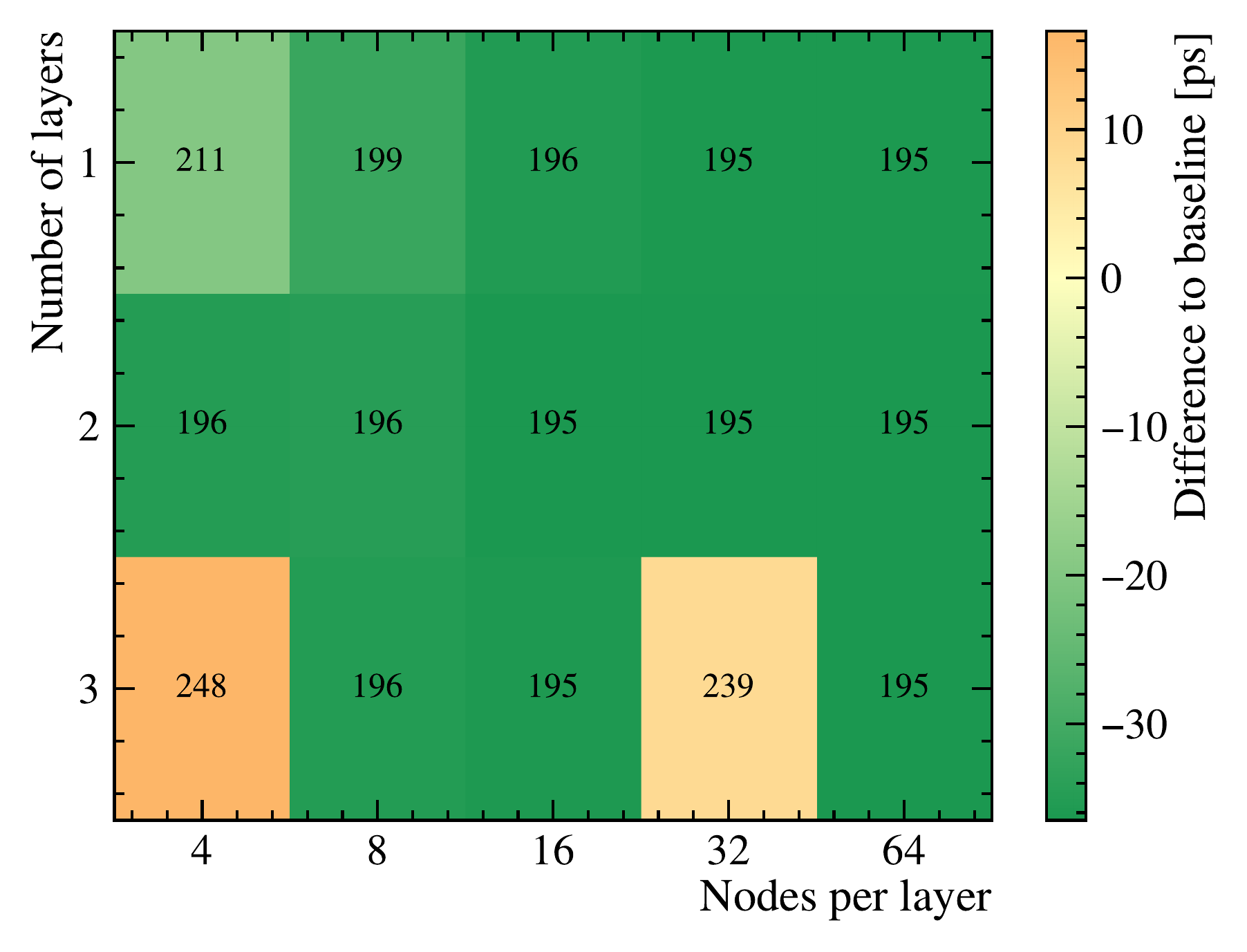}
        \end{subfigure}%
        \hfill%
        \begin{subfigure}{0.5\textwidth}
            \centering
            \includegraphics[width=0.9\textwidth]{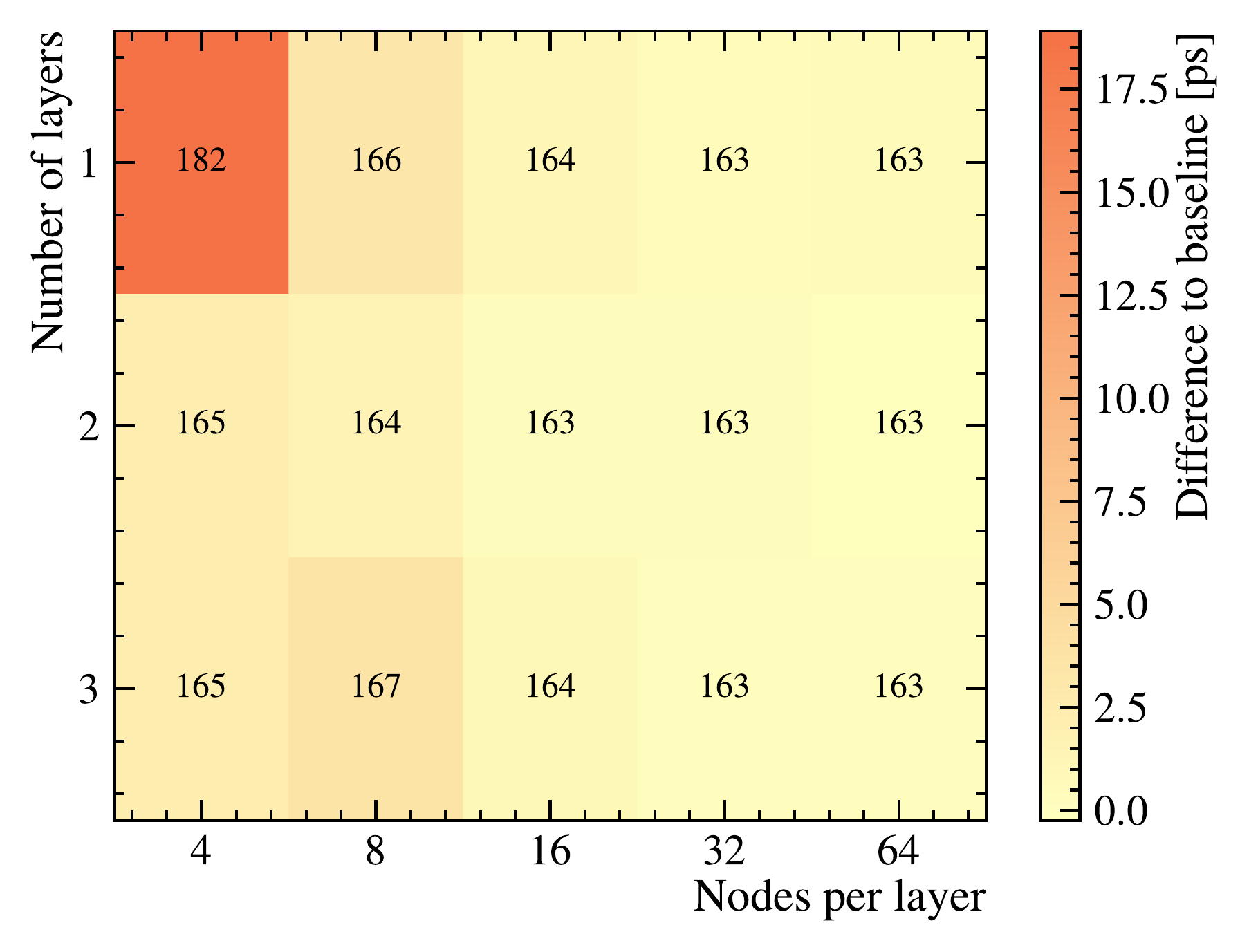}
        \end{subfigure}
        \caption{\qty{100}{\micro\meter} plane}\label{fig:architecture_heatmap:100}
    \end{subfigure}
    \begin{subfigure}{\textwidth}
        \begin{subfigure}{0.5\textwidth}
            \centering
            \includegraphics[width=0.9\textwidth]{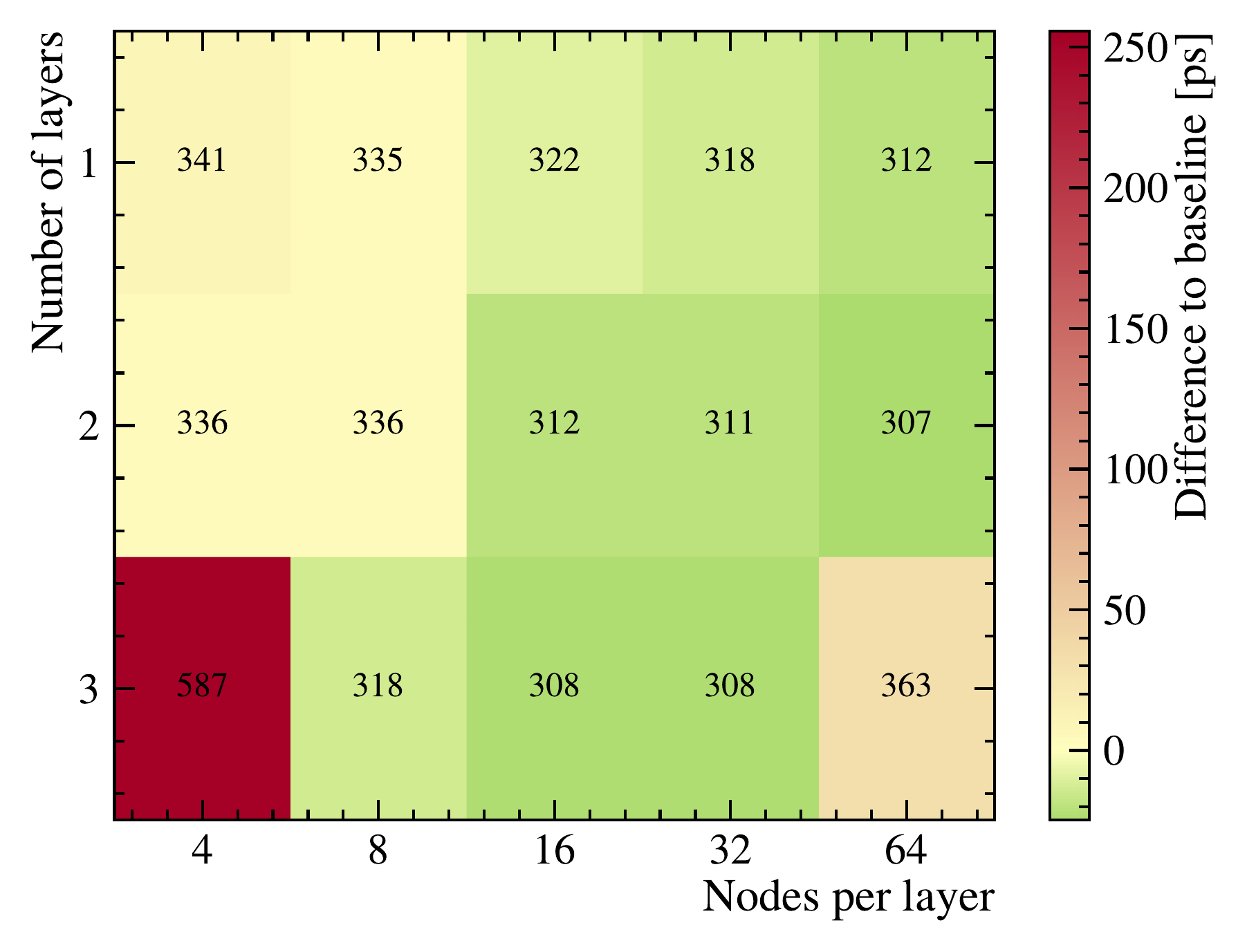}
        \end{subfigure}%
        \hfill%
        \begin{subfigure}{0.5\textwidth}
            \centering
            \includegraphics[width=0.9\textwidth]{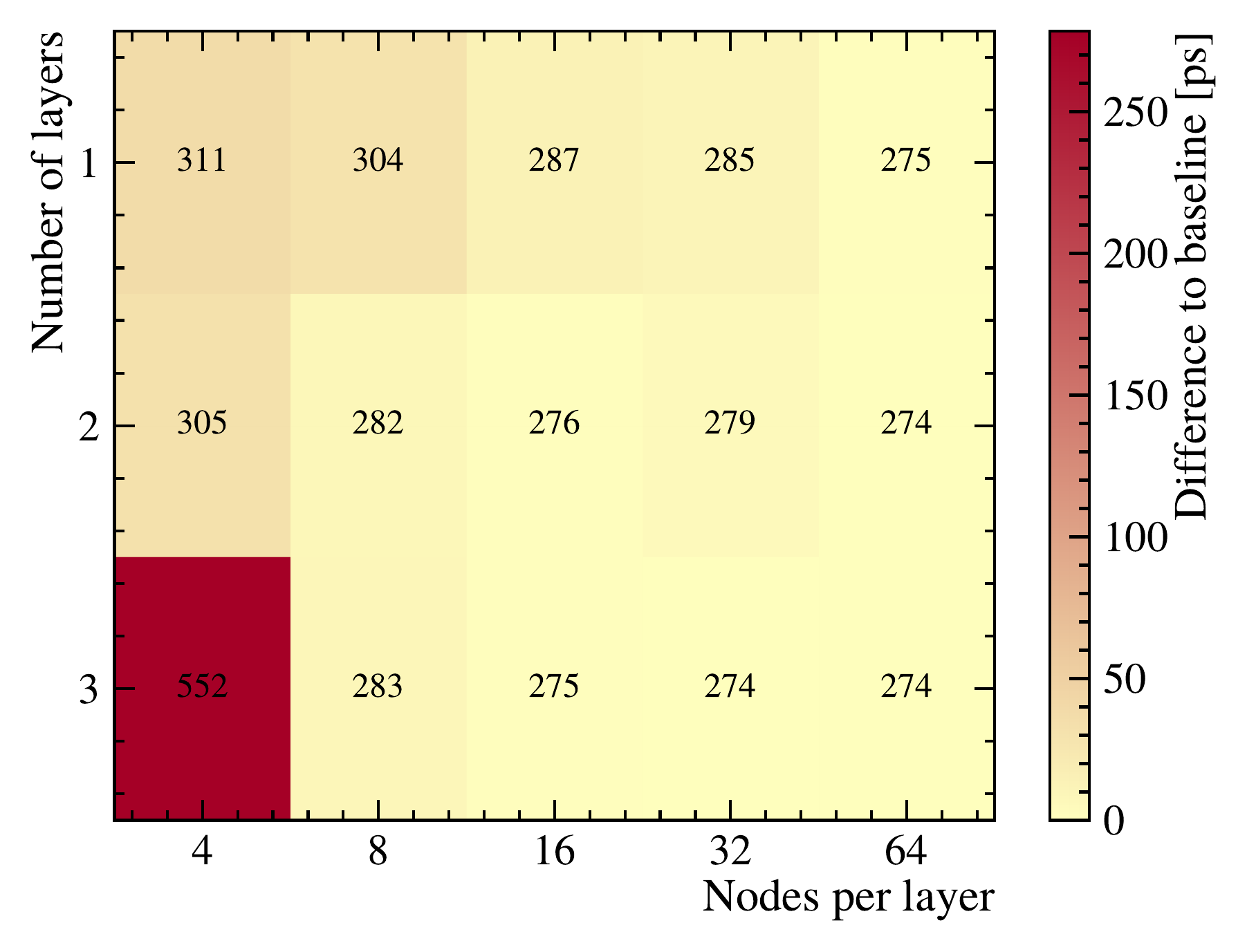}
        \end{subfigure}
        \caption{\qty{300}{\micro\meter} plane}\label{fig:architecture_heatmap:300}
    \end{subfigure}
    \caption{Time resolution as a function of the NN architecture for the \qty{100}{\micro\meter} and \qty{300}{\micro\meter} planes with 10 (left) and 5000 (right) hits per pixel available for training.
    The colour scale shows the improvement with respect to the fit-based baseline with the same number of calibration hits per pixel.
    The numbers displayed in each bin are the absolute time resolution in \unit{\pico\second}.
    }\label{fig:architecture_heatmap}
\end{figure}

\Cref{fig:architecture_heatmap} shows the resulting time resolution as a function of width and depth, for both plane thicknesses and for two representative numbers of calibration hits per pixel (10 and 5000).
For the \qty{100}{\micro\meter} planes, almost all architectures converge to a similar resolution and provide a significant improvement over the fit-based baseline when only limited calibration data are available, with only the smallest NNs (1L4N) lacking sufficient capacity.
With limited calibration data, some of the deeper NNs do not reach the performance of the more compact networks.
For example, the 3L4N and 3L32N architectures give a worse resolution in \cref{fig:architecture_heatmap:100} (left).
In the 3L4N case, the network is deep but narrow, leading to unstable gradients during training.
For the 3L32N case, the network is over-parameterised, leading to overfitting.
In both cases, this leads to poor generalisation performance and often premature early stopping due to the validation loss and less robust training.

For the \qty{300}{\micro\meter} planes, the more complex timewalk behaviour demands more capacity, with at least two layers of sixteen neurons (2L16N) typically needed to reach the best resolution.
As for the thinner planes, some of the largest NNs give a worse resolution at low sample sizes.
This is seen for the 3L64N architecture in \cref{fig:architecture_heatmap:300} (left), again due to the network being over-parameterised and thus less robust.

Across both plane thicknesses, the resolution is more sensitive to the NN's width than to its depth.
The three-layer, four-neuron network (3L4N) is generally the worst-performing architecture, indicating that its small width provides insufficient capacity despite its greater depth.
Overall, the 2L16N architecture offers a robust compromise between a good time resolution and consistent performance across the full range of calibration samples sizes, and is therefore chosen as the default for the remainder of this study.

\begin{figure}
    \centering
    \begin{subfigure}{0.5\textwidth}
        \centering
        \includegraphics[width=0.9\textwidth]{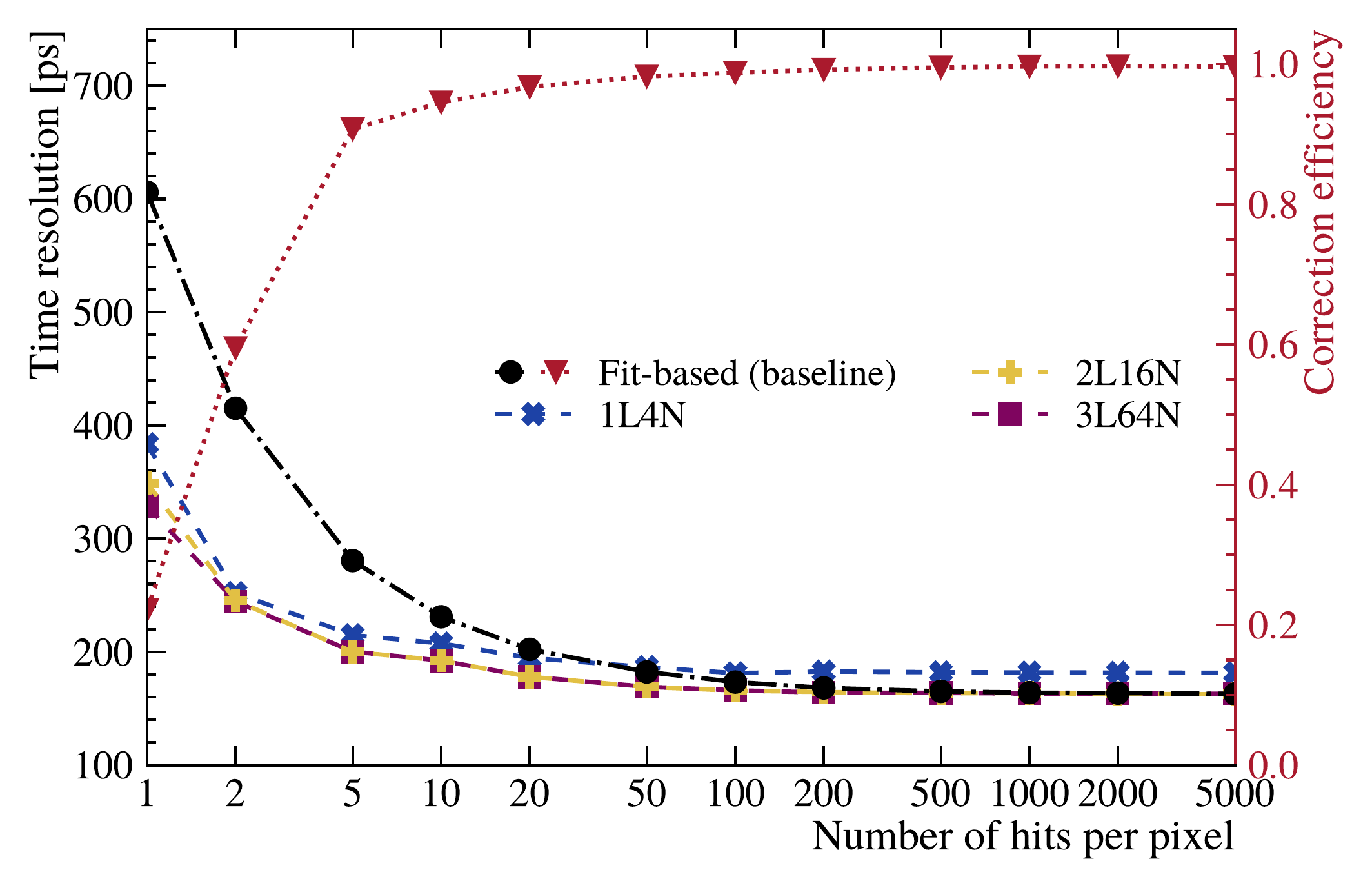}
        \caption{\qty{100}{\micro\meter} plane}\label{fig:result:100}
    \end{subfigure}%
    \hfill%
    \begin{subfigure}{0.5\textwidth}
        \centering
        \includegraphics[width=0.9\textwidth]{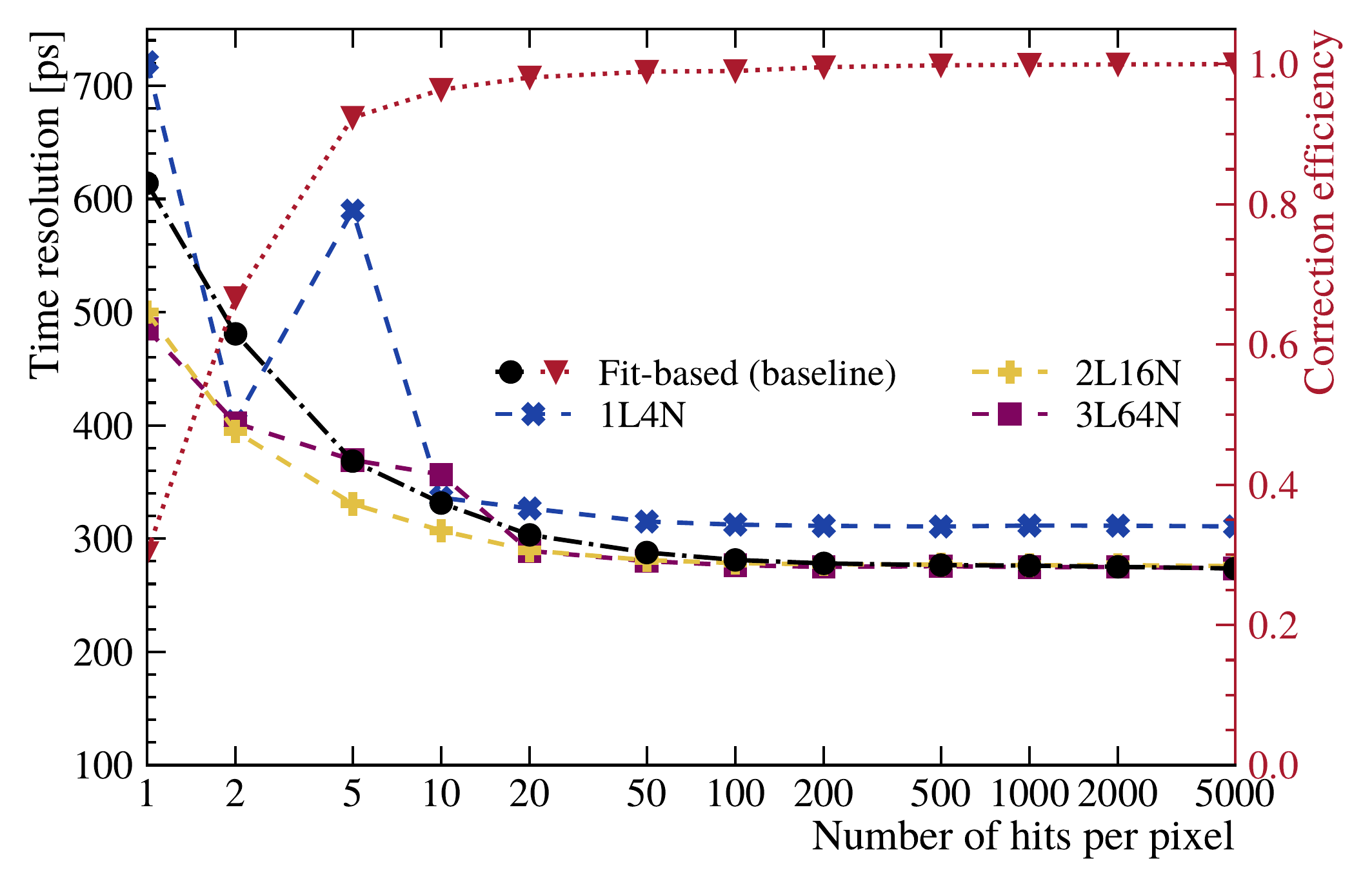}
        \caption{\qty{300}{\micro\meter} plane}\label{fig:result:300}
    \end{subfigure}
    \caption{Time resolution and correction efficiency for clusters as a function of the number of calibration hits per pixel for the fit-based and NN-based timewalk correction methods.
    Three representative NN architectures are shown, denoted 1L4N, 2L16N, and 3L64N.
    }\label{fig:result}
\end{figure}

\Cref{fig:result} compares three representative architectures, 1L4N, 2L16N, and 3L64N, across the full range of calibration sample sizes.
The 1L4N network is too small to capture the timewalk behaviour: it consistently performs worse than the other two, and is even outperformed by the fit-based baseline for large sample sizes.
The 2L16N and 3L64N networks perform similarly, and both consistently outperform the fit-based baseline at low and high sample sizes alike.
At five and ten hits per pixel, however, the 3L64N architecture gives a worse resolution than 2L16N for the \qty{300}{\micro\meter} plane, as shown in \cref{fig:result:300}.

\subsection{Comparison with the fit-based baseline}\label{sec:results:efficiency}
The fit-based and NN-based methods are evaluated using between \num{1} and \num{5000} calibration hits per pixel, with the default 2L16N architecture used for the NN-based correction.
Each calibration sample contains an equal number of hits from all \num{2400} pixels, corresponding to between \num{2400} and \num{12} million hits in total.
Thus, even with the smallest calibration sample size of one hit per pixel, the plane-wide NN and binned fits use \num{2400} hits.
\Cref{fig:result} shows the resulting time resolution and correction efficiency of both methods.

The fit-based procedure requires substantially more calibration data to reach full efficiency.
On the \qty{100}{\micro\meter} planes, its correction efficiency exceeds \qty{99}{\percent} only above \num{500} hits per pixel, and falls to about \qty{20}{\percent} at a single hit per pixel.
The \qty{300}{\micro\meter} planes are less demanding, reaching \qty{99}{\percent} efficiency already at \num{100} hits per pixel.
Because clusters are combined into tracks across at least four, and typically eight, planes, even modest per-plane inefficiencies at low sample sizes can translate into a significant loss of tracks.
The NN does not share this limitation, since its efficiency remains \qty{100}{\percent} regardless of the amount of calibration data available.

The time resolution follows a similar pattern: it decreases as the number of calibration hits per pixel increases, converging to the same asymptotic value for both methods once sufficient data are available.
Part of this overall improvement is common to both methods, since the per-pixel offset corrections in steps~1 and~3 also benefit from the larger number of hits, independent of the timewalk correction itself.
Nevertheless, the NN-based correction retains a clear advantage over the fit-based method when limited calibration data are available: at one hit per pixel, it improves the resolution from \qty{700}{\pico\second} to \qty{350}{\pico\second} for the \qty{100}{\micro\meter} planes, and from \qty{650}{\pico\second} to \qty{550}{\pico\second} for the \qty{300}{\micro\meter} planes.
Above \num{100} hits per pixel, where the bin-by-bin fits are well constrained, both methods converge to about \qty{180}{\pico\second} and \qty{300}{\pico\second}, respectively.

The NN reaches this asymptotic performance already with about \num{100} hits per pixel, whereas the fit-based method requires about \num{500}.
Thus, under the conditions studied, the NN requires only about \qty{20}{\percent} of the calibration data to reach the same performance, allowing the required calibration sample to be acquired five times faster.
This is particularly important when collecting sufficient samples sizes is difficult, for example when the beam intensity or beam size is limited.

The result for the \qty{300}{\micro\meter} planes also illustrates the importance of including all relevant input variables.
In Ref.~\cite{Akiba:2025flo}, a correction based only on charge achieves a resolution of approximately \qty{500}{\pico\second}, while additionally accounting for the intrapixel position improves it to below \qty{300}{\pico\second}.
For the NN, incorporating this information only requires adding the position coordinates as inputs, allowing their correlations with charge to be learned directly.
For the fit-based method, the same extension requires choosing a binning in both position coordinates and performing a separate fit to the measured charge in every sufficiently populated bin.
The timewalk behaviour may also depend on further variables, such as the track angle in situations where the incidence angle varies, or charge information from neighbouring pixels.
Extending the fit-based method to account for these dependencies would require additional, sufficiently populated binning dimensions, whereas the NN developed here can incorporate them directly as inputs in planned future studies.

\section{Conclusion}\label{sec:conclusion}
This paper presented a NN-based timewalk correction for the \timepixfour readout chip using a compact feed-forward NN replacing the conventional bin-by-bin fits.
Its principal methodological advantage is that it learns the correction directly from the input variables, avoiding both an explicit functional ansatz and a binning choice.

A scan over the NN width and depth shows that the resolution improves with increasing model complexity, although some of the largest networks give a worse resolution at selected small calibration sample sizes, due to less robust training when data are scarce.
A two-layer, sixteen-neuron network offers a robust compromise across this trade-off.

The NN is compared against the fit-based baseline using data from the \timepixfour beam telescope collected at the \cern SPS H8 beamline with a \qty{180}{\giga\electronvolt\per\clight} mixed hadron beam in 2025.
Both methods calculate the timewalk correction from the measured charge and the reconstructed intrapixel impact position.
The two methods reach the same asymptotic time resolution, about \qty{180}{\pico\second} and \qty{300}{\pico\second} for the \qty{100}{\micro\meter} and \qty{300}{\micro\meter} sensors, but the NN reaches this performance with only \qty{20}{\percent} of the calibration data required by the fit-based method.
Below \num{100} hits per pixel, the NN degrades far more gracefully.
The fit-based method also loses correction efficiency at small calibration sample sizes, falling to about \qty{20}{\percent} on the \qty{100}{\micro\meter} sensors at a single hit per pixel.
The NN, by contrast, retains full efficiency throughout.

The present study deliberately uses only the measured charge and intrapixel impact position, matching the inputs of the fit-based correction to provide a fair comparison of the two methods.
Future studies are planned to investigate whether information from neighbouring pixels provides further improvements, and if the approach can be extended to take into account different track incidence angles.
These variables can be included as additional NN inputs without introducing further explicit binning dimensions.
However, their effect on the required calibration sample size and the robustness of the correction remains to be evaluated.

The fivefold reduction in the data needed to reach the asymptotic resolution allows calibration data to be acquired more quickly.
The NN approach is therefore particularly promising when acquiring sufficient calibration data is difficult due to insufficient beam intensity or size.

\section*{Acknowledgements}
We would like to express our gratitude to our colleagues in the \cern accelerator departments for the excellent performance of the beam in the SPS North Area. 
We thank the \lhcb testbeam coordinator, Loris Martinazzoli, for the support and coordination of the testbeam campaigns.
We acknowledge the ASIC designers of \timepixfour and PicoTDC, and the designers of the SPIDR4 readout system.
We gratefully acknowledge the computing resources and batch services provided by \cern.
We also thank the open-source software communities that supported this work.
We acknowledge the \cern EP R\&D program's contribution to the construction of the \timepixfour beam telescope.
We also gratefully acknowledge the support from the Netherlands Organisation for Scientific Research (NWO),
and the Science and Technology Facilities Council (UK, grant number ST/V003151/1).

\clearpage
\addcontentsline{toc}{section}
{References}
\setboolean{inbibliography}{true}

\bibliographystyle{JHEP}
\bibliography{main}
\end{document}